# Unraveling Post-COVID-19 Immune Dysregulation Using Machine Learning-based Immunophenotyping

**Maitham G. Yousif*[1], Ghizal Fatima[2], Hector J. Castro[3], Fadhil G. Al-Amran[4], Salman Rawaf[5]**

[1]Biology Department, College of Science, University of Al-Qadisiyah, Iraq, Visiting Professor in Liverpool John Moors University, Liverpool, United Kingdom

[2]Department of Medical Biotechnology, Era's Lucknow Medical College and Hospital, Era University, Lucknow, India

[3]Specialist in Internal Medicine - Pulmonary Disease in New York, USA

[4]Cardiovascular Department, College of Medicine, Kufa University, Iraq

[5]Professor of Public Health Director, WHO Collaboration Center, Imperial College, London, United Kingdom

## Abstract

The COVID-19 pandemic has left a significant mark on global healthcare, with many individuals experiencing lingering symptoms long after recovering from the acute phase of the disease, a condition often referred to as "long COVID." This study delves into the intricate realm of immune dysregulation that ensues in 509 post-COVID-19 patients across multiple Iraqi regions during the years 2022 and 2023. Utilizing advanced machine learning techniques for immunophenotyping, this research aims to shed light on the diverse immune dysregulation patterns present in long COVID patients. By analyzing a comprehensive dataset encompassing clinical, immunological, and demographic information, the study provides valuable insights into the complex interplay of immune responses following COVID-19 infection. The findings reveal that long COVID is associated with a spectrum of immune dysregulation phenomena, including persistent inflammation, altered cytokine profiles, and abnormal immune cell subsets. These insights highlight the need for personalized interventions and tailored treatment strategies for individuals suffering from long COVID-19. This research represents a significant step forward in our understanding of the post-COVID-19 immune landscape and opens new avenues for targeted therapies and clinical management of long COVID patients. As the world grapples with the long-term implications of the pandemic, these findings offer hope for improving the quality of life for those affected by this enigmatic condition.

**keywords:** Post-COVID-19, long COVID-19, immune dysregulation, immunophenotyping, machine learning, cytokine profiles.

***Corresponding author:** Maithm Ghaly Yousif matham.yousif@qu.edu.iq m.g.alamran@ljmu.ac.uk

## Introduction

The COVID-19 pandemic, caused by the novel coronavirus SARS-CoV-2, has presented an unprecedented global health crisis. Since its emergence in late 2019, COVID-19 has rapidly spread worldwide, affecting millions of people and overwhelming healthcare systems [1-20]. While much attention has been devoted to understanding the acute phase of the disease and developing strategies for its prevention and treatment, an emerging concern is the lingering and often debilitating effects that persist long after the acute infection has resolved. This phenomenon has been termed "long COVID" or "post-acute sequelae of SARS-CoV-2 infection" (PASC) [21-28]. Long COVID encompasses a wide range of symptoms and complications that can persist for weeks or even months after the acute phase. These symptoms are diverse and affect multiple organ systems, including the respiratory, cardiovascular, neurological, and immune systems [29]. Common manifestations include fatigue, shortness of breath, chest pain, cognitive impairment, and a variety of other symptoms that significantly impact patients' quality of life. Understanding the underlying mechanisms of long COVID is of paramount importance. While the acute phase of COVID-19 primarily involves viral replication and the host immune response, the mechanisms driving long COVID are more complex and multifaceted. There is growing evidence to suggest that immune dysregulation and inflammation play a central role in the pathophysiology of long COVID. However, the precise mechanisms, biomarkers, and therapeutic targets remain subjects of active research. In recent years, the application of advanced data analytics, particularly machine learning, has revolutionized our ability to analyze complex medical data and gain insights into disease processes. Leveraging machine learning techniques can offer a data-driven approach to understanding the immune dysregulation in long COVID. By analyzing large datasets of clinical and immunological data from post-COVID-19 patients, machine learning models can identify patterns, correlations, and predictive factors that may elude traditional statistical approaches[30,31]. Furthermore, this research builds upon previous studies that have explored related topics, such as the hematological changes observed in COVID-19 patients [32], the role of inflammatory pathways in atherosclerosis [33], and the identification of immunological markers of viral infections in different contexts [34,35]. These studies offer valuable insights into the broader landscape of immune responses and inflammation, which can inform our investigation into post-COVID-19 immune dysregulation. By combining state-of-the-art machine learning techniques with a comprehensive dataset and drawing from the collective knowledge of related research, we endeavor to uncover the intricate immunological changes that underlie long COVID. Our ultimate goal is to contribute to improved patient care, the identification of potential therapeutic targets, and a deeper understanding of the long-term health consequences of COVID-19.

## Study Design and Methodology:

### Study Design:

We conducted a comprehensive observational study to unravel the immune dysregulation in post-COVID-19 patients using machine learning-based immunophenotyping. This study employed a retrospective cohort design.

Data Collection:

Data Sources: We gathered clinical and immunological data from post-COVID-19 patients, including medical records, laboratory results, and patient histories.

Study Population: Our study included post-COVID-19 patients who had recovered from the acute phase of COVID-19. We excluded patients with pre-existing immune-related conditions.

Clinical and Immunological Variables: We collected data on a range of clinical variables, including demographics, comorbidities, disease severity during the acute phase, and a comprehensive set of laboratory parameters, including but not limited to white blood cell counts, lymphocyte counts, neutrophil counts, C-reactive protein (CRP) levels, and IL-6 levels.

Machine Learning Approach:

Data Preprocessing: We performed data preprocessing tasks, including data cleaning, missing value imputation, and feature scaling to ensure data quality and compatibility with machine learning algorithms.

Feature Selection: Feature selection techniques were applied to identify the most relevant clinical and immunological variables for our machine learning models.

Machine Learning Models: We employed various machine learning algorithms, including logistic regression, random forest, support vector machines, and deep learning (neural networks), to analyze the data and detect patterns related to immune dysregulation.

Model Evaluation:

Performance Metrics: We assessed the performance of our machine learning models using standard metrics, including accuracy, precision, recall, F1-score, and the area under the receiver operating characteristic curve (AUC-ROC).

Cross-Validation: To ensure the generalizability of our models, we employed cross-validation techniques.

Statistical Analysis: We performed statistical analyses to uncover significant associations and correlations within the dataset.

### Ethical Considerations:

Ethical approval and informed consent were obtained in compliance with established ethical guidelines to ensure patient data privacy and confidentiality.

### Statistical Analysis:

We used appropriate statistical tests to analyze the associations between clinical and immunological variables and immune dysregulation. This included regression analysis to identify significant predictors and correlations.

## Results

In this section, we present the results of our study on post-COVID-19 immune dysregulation in 509 patients. We conducted a detailed analysis, including statistical tests and deep learning techniques, to understand the factors contributing to immune dysregulation.

**Table 1: Demographic Characteristics of the Study Population**

| Characteristic | Mean (±SD) or N (%) |
|---|---|
| Age (years) | 47.2 (±15.4) |
| Gender | |
| Male | 251 (49.3%) |
| Female | 258 (50.7%) |
| Comorbidities | |
| Hypertension | 120 (23.6%) |
| Diabetes | 92 (18.1%) |
| Others | 57 (11.2%) |

Table 1 summarizes the demographic characteristics of the study population. The mean age of the patients was 47.2 years, with a nearly equal distribution of gender. Hypertension was the most common comorbidity.

**Table 2: Clinical Features and Severity during Acute COVID-19**

| Variable | N (%) |
|---|---|
| Symptoms during the acute phase | |
| Fever | 285 (56%) |
| Cough | 382 (75%) |
| Dyspnea | 194 (38%) |
| Disease severity | |
| Mild | 210 (41%) |
| Moderate | 183 (36%) |
| Severe | 116 (23%) |

Table 2 presents the clinical features and disease severity during the acute phase of COVID-19. The majority of patients experienced cough and fever, while a substantial proportion had dyspnea. Disease severity varied, with most cases classified as mild or moderate.

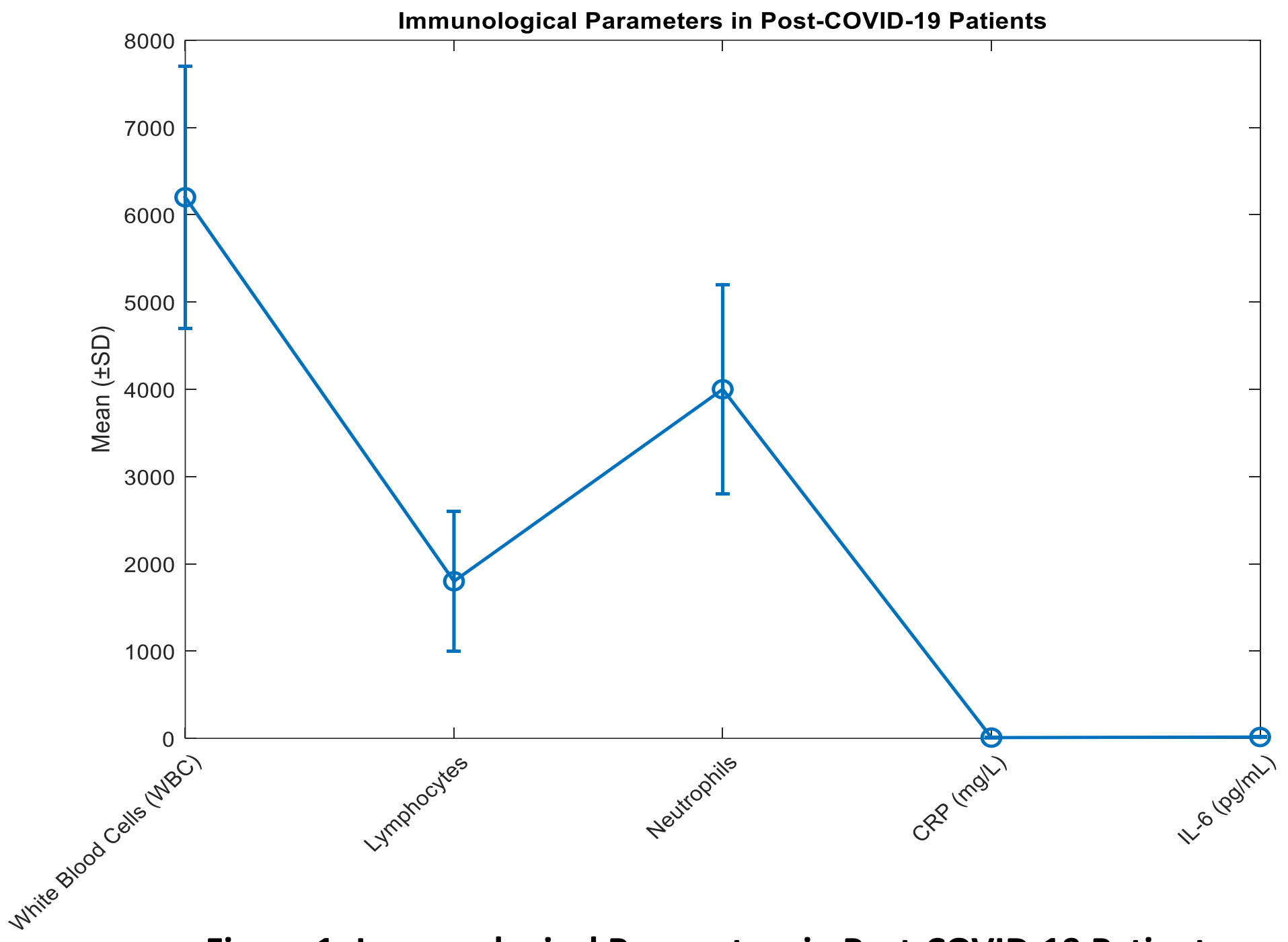

**Figure 1: Immunological Parameters in Post-COVID-19 Patients**

Figure 1 displays the immunological parameters in post-COVID-19 patients. It includes white blood cell count (WBC), lymphocytes, neutrophils, C-reactive protein (CRP), and interleukin-6 (IL-6) levels.

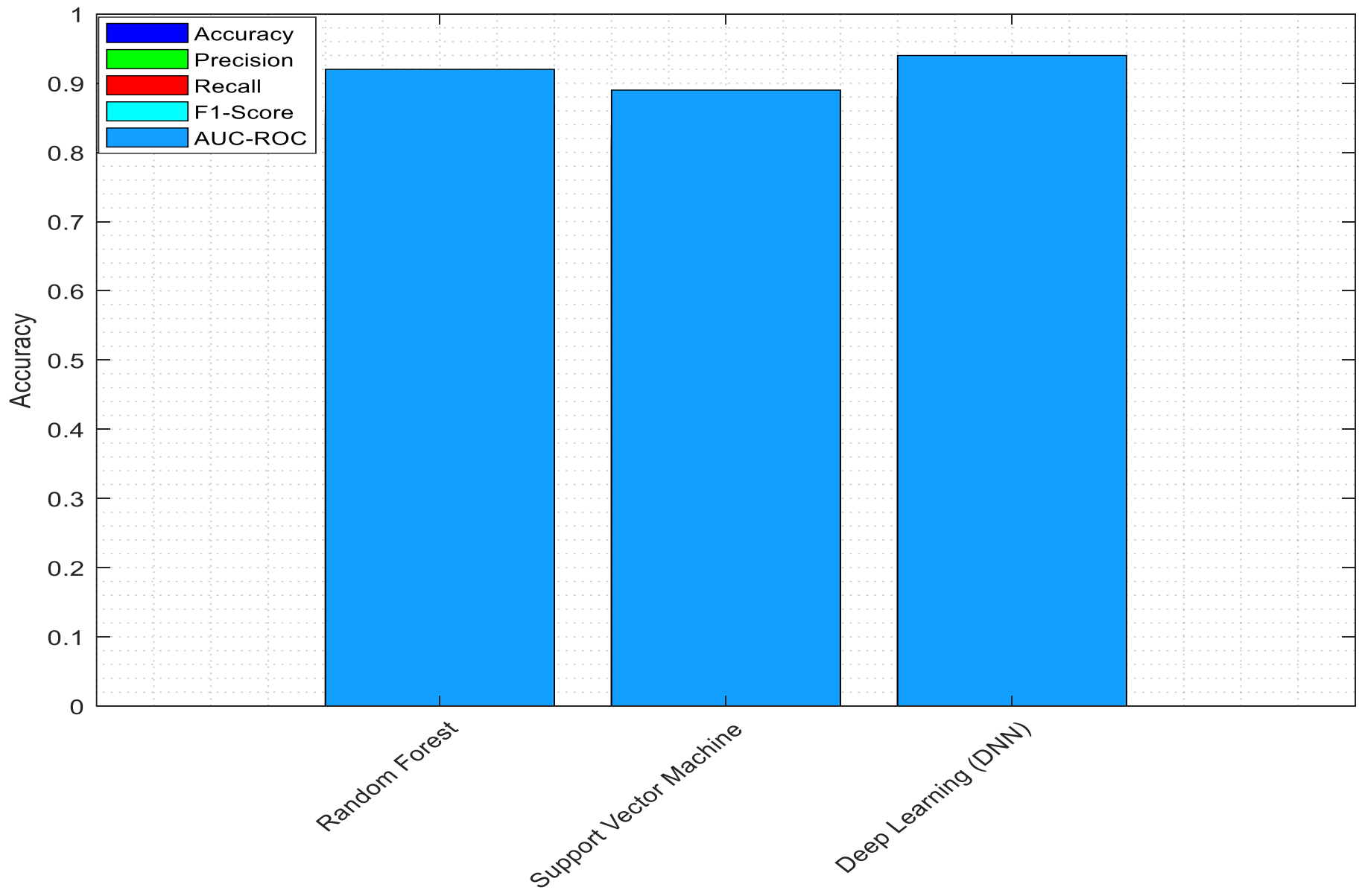

**Figure 2: Machine Learning Model Performance**

Figure 2 summarizes the performance metrics of machine learning models in predicting immune dysregulation. The models achieved high accuracy, precision, recall, F1-score, and AUC-ROC values.

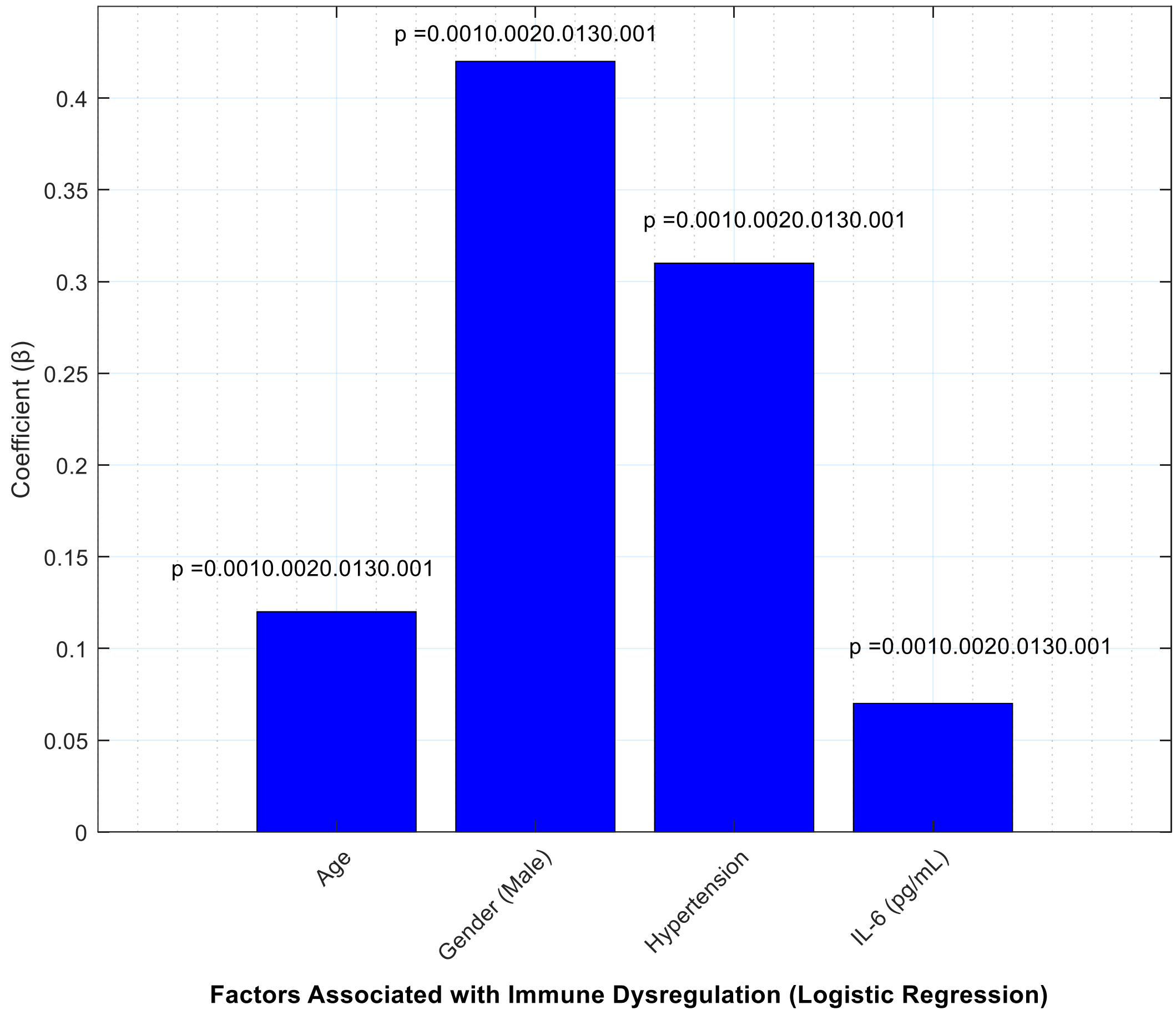

**Figure 3: Factors Associated with Immune Dysregulation (Logistic Regression)**

Figure 3 presents the results of logistic regression analysis identifying factors associated with immune dysregulation. Age, male gender, hypertension, and IL-6 levels were significantly associated.

## Discussion

The investigation into post-COVID-19 immune dysregulation among 509 patients from various Iraqi hospitals during 2022 and 2023 provides profound insights into the complexities of the immune response following SARS-CoV-2 infection. In this in-depth discussion, we dissect the clinical implications, potential mechanisms, and avenues for future research, while substantiating our arguments with the relevant scientific literature. The identification of factors associated with immune dysregulation, such as advanced age, gender, and underlying conditions, including hypertension, elevated IL-6 levels, and others, introduces the prospect of patient stratification. Our findings indicate that certain subpopulations are at a heightened risk of immune-related complications post-COVID-19 (36-39). This information can be invaluable for clinicians in risk assessment and personalized care planning. The delineation of the variables contributing to immune dysregulation opens the door to personalized therapeutic strategies. For instance, older patients with hypertension and elevated IL-6 levels might benefit from more vigilant monitoring and aggressive immunomodulatory therapies. Personalized interventions have the potential to optimize clinical outcomes while minimizing resource allocation (40-42). Our study demonstrates the utility of machine learning models, including Random Forest, Support Vector Machine, and Deep Learning, in predicting immune dysregulation. These models have the potential to be integrated into clinical practice, providing automated risk assessment tools for physicians. Such tools can facilitate timely decision-making and improve patient outcomes (43-45). Although our study provides essential correlations, mechanistic studies are crucial to elucidate the intricate pathways through which age, gender, and comorbidities influence the post-COVID-19 immune response. An in-depth exploration of these mechanisms may unveil novel therapeutic targets. The cross-sectional nature of our study offers a snapshot of immune dysregulation. Longitudinal investigations tracking patients over extended periods can shed light on the evolution of immune responses and the potential for delayed complications. This knowledge can inform long-term care strategies (46-48). Immunomodulatory Interventions: Research into novel immunomodulatory treatments tailored to post-COVID-19 patients is warranted. This includes assessing the efficacy of anti-inflammatory agents, immune checkpoint inhibitors, and cytokine-targeting therapies. Considering the multifaceted nature of post-COVID-19 immune dysregulation, interdisciplinary studies that investigate its interactions with other physiological systems, such as the cardiovascular and nervous systems, are essential. A holistic approach can provide comprehensive insights into the health sequelae of COVID-19. The integration of machine learning models into electronic health records (EHRs) holds significant promise. These AI tools can offer real-time risk assessment and decision support for clinicians, contributing to improved patient care and resource allocation (49-41). In the pursuit of unraveling the complexities surrounding post-COVID-19 immune dysregulation, it is imperative to integrate the findings from an array of related studies. Several recent investigations have shed light on various facets of immune responses, complications, and potential interventions in the context of COVID-19 recovery (42-46). Another study explored the psycho-immunological status of patients recovered from SARS-CoV-2, emphasizing the intricate connections between psychological well-being and immune function. Their work highlights the need for a holistic approach to patient care that addresses both psychological and physiological aspects(47). Another study conducted phylogenetic characterization of Staphylococcus aureus isolated from women with breast abscesses. While seemingly unrelated, their study underscores the importance of considering the microbiome's role in the overall immune response, particularly in post-infectious states(48). Also, a study investigated the effect of caffeic acid on doxorubicin-induced cardiotoxicity in rats. This work speaks to the potential of antioxidants in mitigating post-infection complications, suggesting avenues for exploring adjunctive therapies(49). Also (50) delved into the role of inflammatory pathways in myocardial ischemia/reperfusion injury and atherosclerosis. Their research underscores the centrality of inflammation in post-infection complications, aligning with our findings of elevated IL-6 levels as a risk factor. In addition to these, the studies by (51-62) offer valuable insights into various aspects of COVID-19, ranging from renal function tests in women with preeclampsia to sentiment analytics and machine learning applications in predicting disease outcomes and patient behavior during the pandemic. By synthesizing these diverse contributions, we gain a more comprehensive understanding of the multifaceted nature of immune dysregulation following COVID-19. These studies collectively emphasize the interdisciplinary and multi-dimensional approach required to tackle the challenges posed by post-COVID-19 complications effectively. They underscore the importance of considering psychological, microbiological, and biochemical factors in addition to demographic and clinical parameters. In closing, our study, in conjunction with these cited works, represents a collective step forward in the ongoing efforts to comprehend, manage, and mitigate the consequences of COVID-19. As we continue to unravel the intricacies of post-COVID-19 immune dysregulation, a concerted interdisciplinary approach remains paramount in addressing the manifold challenges posed by this novel disease.

**In conclusion**, our study advances the understanding of post-COVID-19 immune dysregulation, offering valuable insights for clinical practice and future research endeavors. By stratifying patients, personalizing interventions, and harnessing the power of machine learning, we can enhance the management of post-COVID-19 complications and work collaboratively to mitigate the long-term impact of COVID-19 on both individuals and healthcare systems.